\documentclass[superscriptaddress,amssymb,twocolumn,aps]{revtex4}

\usepackage{graphicx,amsmath,color}
\usepackage{booktabs,array,dcolumn}
\usepackage{soul}

\newcommand{\eq}[1]{Eq.~(\ref{#1})}

\newcommand{\fig}[1]{Fig.~\ref{#1}}

\newcommand{\bn}{ {\bf n} }
\newcommand{\be}{ {\bf e} }
\newcommand{\ba}{ {\bf m} }
\newcommand{\bb}{ {\bf l} }

\newcommand{\bm}{ {\bf m} }

\newcommand{\bhu}{ {\bf u} }

\newcommand{\pp}{{\mathcal P}_{2}}
\newcommand{\eeq}{ \end{equation} }
\newcommand{\beq}{ \begin{equation} }
\newcommand{\eea}{ \end{eqnarray} }
\newcommand{\bea}{ \begin{eqnarray} }

\begin{document}

\title{Nemato-elasticity of hybrid molecular-colloidal liquid crystals}

\author{B. Senyuk}
\affiliation{Department of Physics, University of Colorado, Boulder, CO, USA}
\author{H. Mundoor}
\affiliation{Department of Physics, University of Colorado, Boulder, CO, USA}
\author{I. I. Smalyukh}
\affiliation{Department of Physics, University of Colorado, Boulder, CO, USA}
\affiliation{Department of Electrical, Computer, and Energy Engineering, Materials Science and Engineering Program and Soft Materials Research Center, University of Colorado, Boulder, CO, USA}
\affiliation{Chemical Physics Program, Departments of Chemistry and Physics, University of Colorado, Boulder, CO, USA}
\affiliation{Renewable and Sustainable Energy Institute, National Renewable Energy Laboratory and University of Colorado, Boulder, CO 80309, USA}
\author{H. H. Wensink}
\affiliation{Laboratoire de Physique des Solides, Universit\'e Paris-Saclay  \& CNRS, UMR 8502, 91405 Orsay, France}
 
\email{rik.wensink@universite-paris-saclay.fr}

\date{\today}

\begin{abstract}

Colloidal rods immersed in a thermotropic liquid-crystalline solvent are at the basis of so-called hybrid liquid crystals which are characterized by tunable  nematic fluidity with symmetries ranging from conventional uniaxial nematic or anti-nematic  to  orthorhombic [Mundoor {
\em et al.}, {\em Science} {\bf 360}, 768 (2018)].   We  provide a theoretical analysis of the elastic moduli of such systems by considering   interactions between the individual rods with the embedding solvent through surface-anchoring forces, as well as steric and electrostatic interactions between the rods themselves.  For uniaxial systems the presence of  colloidal rods generates a marked increase of the splay elasticity  which we found to be in quantitative agreement with experimental measurements. For orthorhombic hybrid liquid crystals we provide estimates of all twelve elastic moduli and show that only a small subset of those elastic constants play a relevant role in describing the nemato-elastic properties. The complexity and possibilities related to identifying the elastic moduli in experiments are briefly discussed.  The results are expected to be helpful in stimulating future studies on defect-dressed ``topological" colloids and the analysis of boundary-driven phenomena where the non-trivial director topology generated by biaxial nematics plays a key role.

\end{abstract}

\maketitle

\section{Introduction}

Liquid crystalline (LC) phases are found in a large variety of material systems, including the classic examples of anisotropic small molecules and colloidal particles like rods and discs \cite{degennes1993,chaikin2000principles}. Thermotropic nematic LC phases formed by molecular rods within a chemically homogeneous medium, each about a nanometer in length, are the most widely known because of their widespread use in displays and electro-optic devices, where they are stable in a broad temperature range in-between crystalline and isotropic fluid phases. Lyotropic nematic LCs formed by colloidal rod-like particles suspended in a fluid host medium, like water, constitute another classic example of such a nematic mesophase,  although the  physics behind its formation is different from  that of thermotropic nematics \cite{Onsager}. Nematic colloids, where a thermotropic LC is used to host colloidal particles being one or several orders of magnitude larger than the molecules of the host medium, have attracted a great deal of interest over the past decades \cite{brochard1970,poulin1997,burylov1994,ramaswamy1996,ruhwandl1997,lubensky1998,loudet2000,gu2000,stark2001,loudet2001,lev2002,musevic2006sci,musevic2008,smalyukh2018,tovkach2012,nych2013,blanc2013,tkalec2013,musevic2013,lagerwall2016,musevis2017book,pergamenshchik2011}. While spherical colloidal inclusions  are the most widely studied, particles with various geometrically and topologically complex shapes immersed in a LC host have been studied too \cite{lapointe2009,lapointe2010,pergamenshchik2011,liu2014,tkalec2011,zhang2015,lopezleon2011,qliu2013,machon2013,martinez2014,ravnik2015,senyuk2015,hashemi2017,martinez2015,seyednejad2018,pusovnik2019}, including colloidal objects like gold nanorods \cite{koenig2007,koenig2009,liu2010,qi2011,evans2011,engstrom2011,qliu2012,coursault2012,senyuk2012,choudhary2014,senyuk2013pre}, bacteria \cite{smalyukh2008,agarwal2008,mushenheim2014}, carbon nanotubes \cite{ouldmoussa2013,lynch2002,lagerwall2008,schymura2010,agha2016,yadav2016,draude2021} and graphene sheets \cite{twombly2013,alzangana2016,draude2021}. Although some colloidal inclusions were composed of monodomain magnetic or electric dipoles \cite{brochard1970,shuhsia1983,mertelj2013,hess2015,lisjak2018,fleury2020}, the  majority of nematic colloids studied thus far involved dielectric particles or droplets. The most notable examples relate to  cases in which the orientations of the anisotropic colloidal nanoparticles mimicks  the director pattern of the molecular or surfactant-based nematic host medium \cite{lapointe2009,lapointe2010,alzangana2016,lynch2002,lagerwall2008,schymura2010,agha2016,yadav2016,draude2021,pergamenshchik2011,tkalec2011,lopezleon2011,qliu2013,machon2013,ravnik2015,hashemi2017,seyednejad2018,twombly2013,pusovnik2019,martinez2014,liu2014,zhang2015,senyuk2015,martinez2015}, though, even more interestingly, in some cases  so-called anti-nematic order of colloidal rods in a nematic host medium was observed as well \cite{qliu2012}. Combining colloidal and LC systems not only resulted in new physical properties, such as polarization-dependent surface plasmon resonance properties in dispersions of gold nanorods within LCs \cite{liu2010}, it also led to the recent discovery of new soft condensed matter phases formed by  molecular-colloidal composite systems. Introduced as LCs that combine lyotropic and thermotropic phase behavior, hybrid molecular-colloidal LCs have been studied in recent years as an experimental platform for the creation of low-symmetry orientationally-ordered fluid organizations \cite{liu2016}, including orthorhombic \cite{mundoor2018} or monoclinic nematic LCs \cite{mundoor2021}, and triclinic colloidal crystals \cite{mundoor2016}.

\begin{figure}
\begin{center}
\includegraphics[width= \columnwidth ]{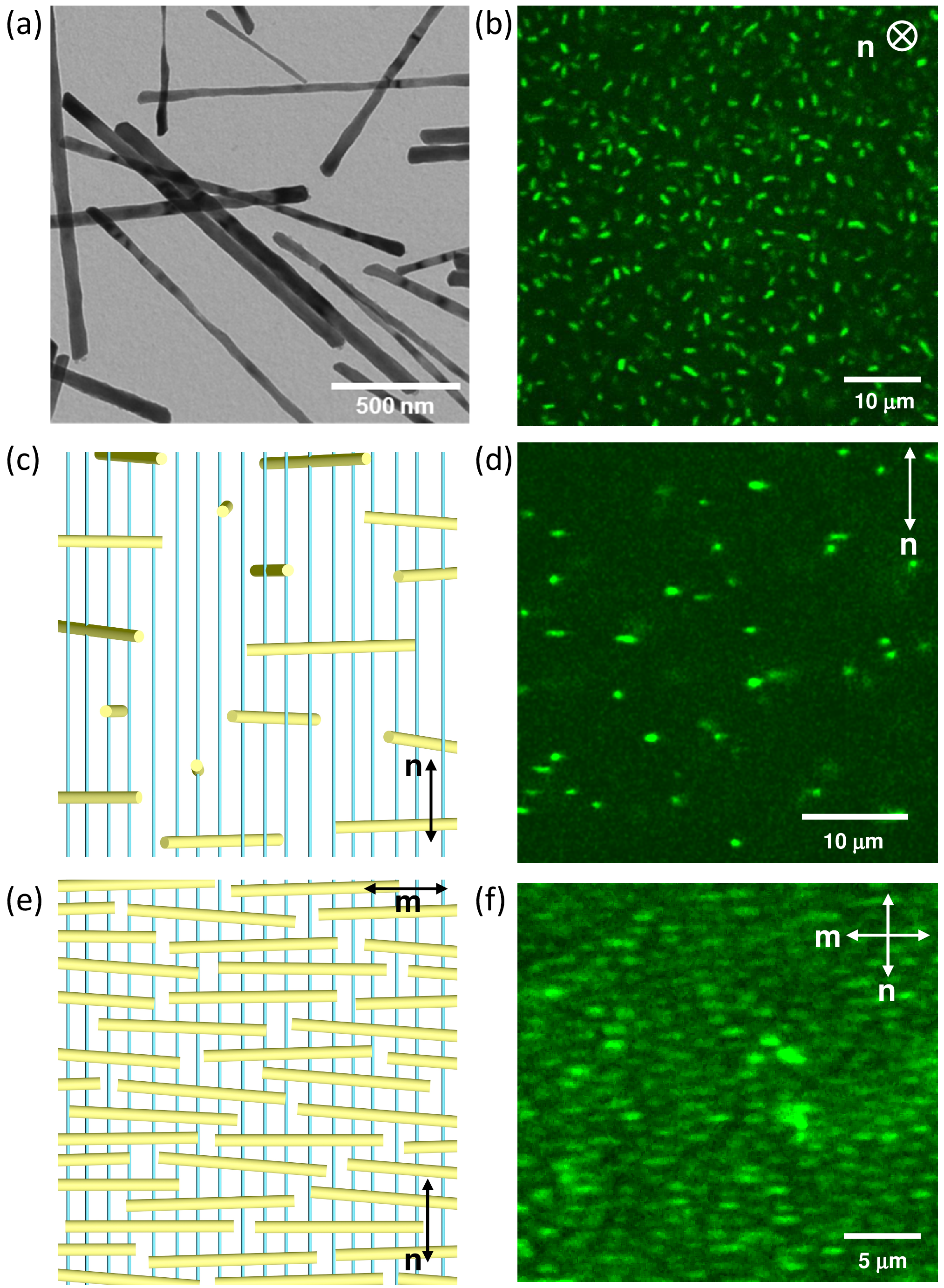}
\caption{ Structure and composition of a hybrid molecular-rod system: (a) Transmission electron micrograph of colloial nanorods after acid treatment. (b, d) Upconversion-based luminescence confocal microscopy images of the nanorods in a uniaxial nematic LC in a homeotropic (b) and planar (d) cell. (c) Schematic overview and  corresponding luminescence confocal microscopy image (d) of the nanorods in a uniaxial nematic LC. (e) Schematic overview and corresponding luminescence confocal microscopy image (f) of the nanorods forming an orthorhombic biaxial nematic LC in a planar cell.  }
\label{exp1}
\end{center}
\end{figure}

Owing to the LC character of the  solvent, surface anchoring forces introduce an orientational coupling between the ordering of solvent molecules  and the colloidal surfaces such that the molecular director tends to favorably align either normal to the colloidal surface (homeotropic anchoring) or tangential to it (planar anchoring) \cite{brochard1970,jerome1991}.   The surface anchoring properties can be tuned by controlling the  properties of the colloidal surfaces \cite{mundoorSA2019}  as well as the  temperature of the LC solvent.  A key attribute of the hybrid LCs is that the surface anchoring energy per particle (relative to its thermal energy) is strong enough for the colloids to experience considerable realigning torques with respect to the solvent director, but at the same time weak enough to avoid bulk disclinations or other topological defects around the colloidal surfaces \cite{stark2001}. These defects possess a well-defined topology and are routinely encountered for relatively large nematic colloidal inclusions with strong surface anchoring. They give rise to strong interparticle forces \cite{poulin1997} leading to dynamically arrested composites \cite{wood2011,stratford2014,musevic2006,yuan2019,yuan2021}.  In contrast, because of weaker elastic distortions and additional electrostatic stabilization, structure formation in hybrid LCs is fully reversible and a rich array of phase transitions can be explored across a wide range of parameters (colloid concentration, temperature, etc.)  without risking the system to be trapped in a non-equilibrium state.

For the case of slender rigid rod-shaped colloids (\fig{exp1}), one can show that homeotropic surface anchoring enforces the rods to align with their main orientation perpendicular to the molecular director. At low rod concentration a uniaxial hybrid LC  is then created in which the rods adopt anti-nematic orientational order [\fig{exp1}(c,d)].

When the concentration of immersed rods is sufficiently large, rod correlations create a biaxial nematic fluid with orthorhombic ($D_{2h}$) point symmetry [\fig{exp1}(e,f)].   The creation of such a well-controlled orthorhombic nematic phase provides strong impetus to revisit its nemato-elastic properties, which is key to understanding the director-field topology that such a low-symmetry nematic generates when exposed to various types of confining boundaries \cite{degennes1993}. Various theoretical routes to consider elasticity of biaxial nematics have been developed \cite{kini1989,singh1994,kapanowski1997,stelzer,longa2, govers1998,stewartbiax} that build upon  the Oseen-Frank theory for uniaxial nematics \cite{oseen1933,frank58}. The principal outcome is that describing the elastic properties of a  biaxial nematic fluid in the absence of chirality requires  twelve elastic constants instead of the three bulk moduli, related to splay, twist and bend fluctuations (see \fig{defo}) for conventional uniaxial nematics. The twelve elastic moduli can, at least formally, be connected to several microscopic features of the constituent particle, for instance, the  biaxial symmetry of the particle which may be of steric origin or emerge from some long-ranged anisotropic dispersive forces \cite{luckhurst2015}.  

The theoretical predictions are, however, difficult to validate without the availability of a well-characterized experimental biaxial nematic system  where such interactions can be modelled with reasonable accuracy and ease. Our hybrid LC meets those criteria given that the immersed rods in view of their large aspect ratio, rigidity and known surface charge properties, closely follow an Onsager-type behavior \cite{Onsager,mundoor2018} while the coupling between the surface anchoring energy and rod orientation can be well understood from a simple mean-field description \cite{liu2016}. A theoretical description based on these two ingredients  allows for a quantitative prediction of the phase behavior over a wide range of rod concentrations and temperatures \cite{mundoor2018}.

In this paper we wish to apply the same modelling strategy to describe the bulk nemato-elasticity of hybrid LCs from a  particle-based theoretical viewpoint.  The discussion broadly falls into two parts. We begin in Sections II and III by considering the case of uniaxial hybrid LCs where the usual splay, bend and twist modes are affected by the presence of anti-nematically oriented colloidal rods,  mostly through surface-anchoring forces \cite{wensinkSM18}. Here, our theoretical predictions are tested against experimental measurements of the splay modulus for low-concentration uniaxial hybrid LCs. We find that our model provides a quantitative prediction of the increase of the splay modulus with the concentration of immersed rods.  

The second part (Section IV) concerns the elastic moduli of the orthorhombic hybrid LC. We demonstrate that the twelve moduli can be effectively classified by considering only their leading order contributions, be they driven by the elasticity of the LC solvent, by surface anchoring \cite{wensinkSM18} or by repulsive interactions between the colloidal rods \cite{straley1973,odijkelastic}.

From our scaling analysis it transpires that there are only a relatively small subset of dominant elastic moduli that can all be connected to the known  elastic moduli of the  nematic solvent, supplemented with weak corrections due to rod-rod correlations. This approach enables us to fully specify the bulk elastic anisotropy of an experimentally realizable biaxial nematic system that holds great potential for exploring  colloidal or granular objects with reconfigurable topology \cite{senyukh2013} and their self-assembly in orthorhombic nematic media. We finish our manuscript by discussing  a number of technical complications associated with measuring the orthorhombic elastic constants in hybrid colloidal-molecular systems (Section V) and propose a roadmap demonstrating the various experimental set-ups that would need to be realized in order to probe all elastic deformations that feature in the continuum elasticity theory for biaxial nematics (Appendix).

 \section{Uniaxial hybrid liquid crystals}

   \begin{figure}
\begin{center}
\includegraphics[width= 1 \columnwidth ]{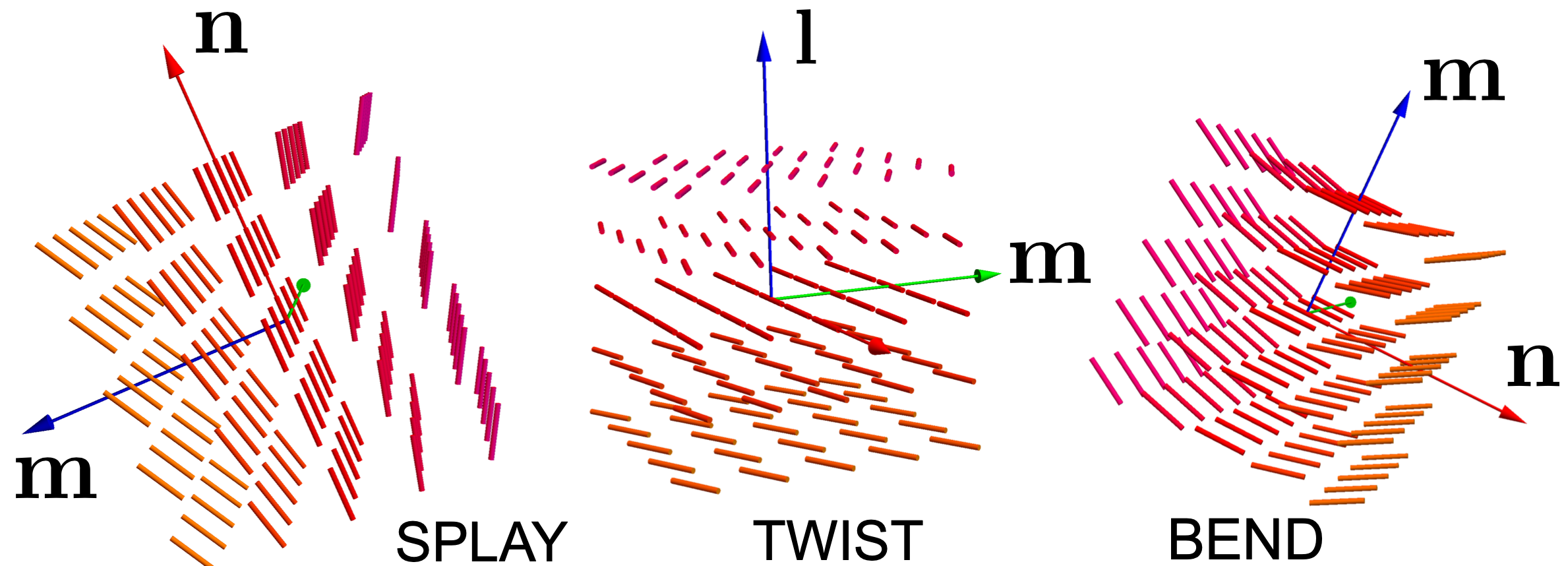}
\caption{ Visualization of the principal director fluctuations $\delta \bn_{m}$ [specified by \eq{eldef1}] of a uniaxial nematic liquid crystal with principal director $\bn$ in a lab frame spanned by the orthonormal tripod $\{ \ba, \bb, \bn \}$.  }
\label{defo}
\end{center}
\end{figure}

Let us define elementary bulk director fluctuations of curvature $\varepsilon  \ll a^{-1}$ (with $a$ the typical molecular size) around the molecular director $\bn$ within a Cartesian director frame $\{ \ba,  \bb, \bn\}$:
\begin{align}
 \delta \bn_{i}^{m} & =  \varepsilon \ba ({\bf r} \cdot\be_{i})
 \label{eldef1}
\end{align}
The deformation vector $\be$ defines the basic distortion pattern:   
\beq
\be_{i} =
\begin{cases}
 \ba , & i=1 \hspace{0.2cm} \text{(splay)} \\
 \bb , & i=2 \hspace{0.2cm} \text{(twist)} \\
  \bn , & i=3 \hspace{0.2cm} \text{(bend)} 
 \end{cases}
\eeq
which define the splay, twist and bend deformations in terms of the position ${\bf r}$ within the  director frame (\fig{defo}). In this study, the focus is on bulk elasticity and we will not consider  surface effects such as  saddle-splay and splay-bend fluctuations which only play a role if the liquid crystal is in contact with a curved surface (see e.g. Ref. \cite{selinger2018} for a recent discussion).

If the system is uniaxial, then there is no preferred orientational order across the plane perpendicular to $\bn$ and one could define a similar set of fluctuation $\delta \bn^{l}_{i}$ upon interchanging the vectors $\ba$ and $\bb$. Both  sets would lead to the same deformation free energy. Clearly, the degeneracy $\delta \bn^{m}_{i}= \delta \bn_{i}^{l}$ no longer holds for the orthorhombic nematic as we will discuss in more detail in Section IV.

We begin by considering a  uniaxial hybrid nematic fluid composed of rods embedded in a uniaxial molecular LC [\fig{exp1}(c,d)]. At low particle concentration,  rod-rod interactions are too insignificant to generate nematic order of the colloidal component alone, and the rods are arranged in an anti-nematic fashion pointing perpendicular to $\bn$ [\fig{exp1}(c)]. Since the hybrid system retains its uniaxial symmetry, only the molecular director matters and its three basic deformation modes (splay, bend and twist) are depicted in \fig{defo}. Elastic restoring forces  are transmitted through the molecular component as well as through interactions between the rod inclusions. 
A straightforward and transparent way to  address the elastic  properties of the hybrid system is to assume a simple superposition of component-specific contributions; one relating to the pure molecular component 5CB (denoted by ``$0$"), a second term accounting for the effect of  surface anchoring (``$s$") and a third contribution arising from steric and electrostatic rod correlations (``$r$"):
\beq
K_{i} \sim K_{i}^{(0)} + K_{i}^{(s)} + K_{i}^{(r)}, \qquad i=1,2,3 
\label{ksuper}
\eeq
At low concentration of rod inclusions we expect the surface-anchoring of the molecular LC at rod surface to play a dominant role in determining the elasticity of the hybrid nematic fluid, with $K_{i}^{(r)} \ll K_{i}^{(s)}$. We now proceed to analyzing each of these contributions in more detail.
 
\subsection{Elasticity generated by surface anchoring }

\begin{figure*}
\begin{center}
\includegraphics[width= 1.6\columnwidth ]{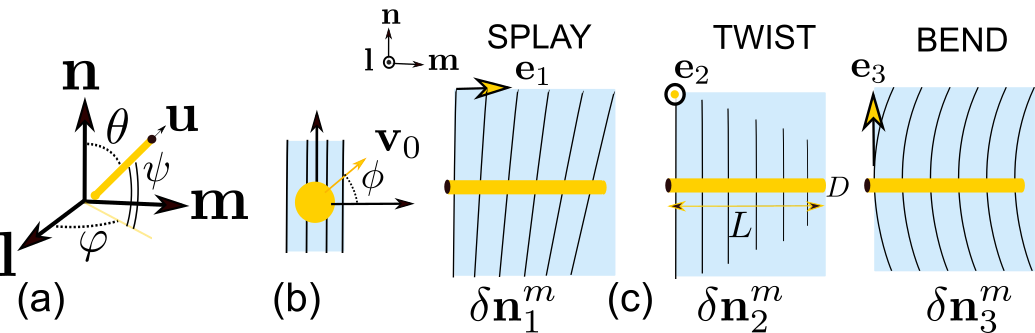}
\caption{  (a) Principal angles describing the orientation $\bhu$ of a single rod with respect to the molecular director $\bn$. (b) Each rod experiences homeotropic surface anchoring across its cylindrical backbone where the molecular director is favored to align along the easy axis ${\bf v}_{0}$ given by a vector parameterizing the circular rod cross-section in terms of the angle $\phi$.  (c)  Illustration of the splay, twist and bend distortions of the molecular director in the presence of a rod-shaped inclusion.   \label{anchorfig}}  
\end{center}
\end{figure*}

Following Ref. \cite{wensinkSM18}, we consider  the Rapini-Papoular surface anchoring free energy \cite{rapini1969} for an ensemble of $N$ cylindrical rods with a length $L$, diameter $D$ in  a volume $V$  and number concentration $\rho = N/V$ immersed in a molecular LC solvent with constant temperature $T$:
\beq
\frac{F_{s}}{V} = -\frac{1}{2} w_{0} \rho \int d \bhu \oint d{\mathcal S} (\bn \cdot  {\bf v}_{0} ({\mathcal S})   )^{2} f_{U}(\bhu \cdot \bn)
\label{fsa}
\eeq
where $w_{0} >0$ denotes the surface anchoring coefficient, $\mathcal S$ the cylindrical rod surface and ${\bf v}_{0}(\phi) = \cos \phi \bhu_{\perp 1} + \sin \phi \bhu_{\perp 2} $ a parameterization for the easy axis normal to the rod surface in terms of an orthonormal rod frame $\{ \bhu, \bhu_{\perp 1} ,\bhu_{\perp 2} \}$. The integral over the rod surface (neglecting end effects for $D \ll L$) can be written as $\int  d{\mathcal S} = (D/2) \int_{0}^{2 \pi} d \phi  \int_{0}^{L} dt $. At fixed polar 
angle $\theta_{e}$ and degenerate azimuthal angle $0 < \varphi \leq 2 \pi$,  the rod orientation probability reads $f_{U}(\bhu)  = \frac{1}{2 \pi} \delta ( \theta - \theta_{e})$ and the surface anchoring energy per particle reads:
\beq
U_{s} = -\frac{\pi}{4} DL w_{0} \sin^{2} \theta_{e} 
\eeq
which  demonstrates that homeotropic anchoring forces the rods to align perpendicular to the molecular director ($\theta_{e} = \pi/2$). The corresponding orientational probability of each rod is given by a Boltzmann 
factor: 
\beq
f_{U}(\theta) \propto  \exp \left [ - \sigma  \pp (\cos \theta) \right ] 
\label{fu}
\eeq
in terms of the anti-nematic field strength $\sigma$ which, at least in the  experimental system at hand, strongly exceeds the thermal energy $k_{B} T$ (with $k_{B}$ Boltzmann's constant):
\beq
\sigma = \frac{\pi LDw_{0}}{6 k_{B}T} \gg 1
\eeq
The fact that surface anchoring generates a very strong re-aligning potential justifies the use of a simpler Gaussian distribution describing small fluctuations of the  meridional angle $\psi = \pi/2 - \theta$
\beq
f_{U}(\psi )  \sim \sqrt{\frac{3\sigma}{(2 \pi )^{3} }} \exp \left  (  -\frac{3}{2} \sigma \psi^{2} \right )
\eeq
which is normalized on the unit sphere via $\lim_{\sigma \rightarrow \infty} \int_{-\infty}^{\infty} d \psi \int_{0}^{2 \pi} d \varphi f_{U}(\psi) = 1$. 

The surface anchoring free energy is easily  generalized to the case of (weakly) non-uniform molecular director fields $\bn + \delta \bn_{i}^{m}$ given by \eq{eldef1}.
We expand \eq{fsa} up to  ${\mathcal O} (\varepsilon^{2})$ and write down the free energy cost associated with infinitely weak long-wavelength molecular director deformations.  The associated elastic contributions originating from surface anchoring are then expressed as: 
\begin{align}
K_{i}^{(s)} = & \frac{\delta F_{s}}{\frac{1}{2} \varepsilon^{2} V} =-\frac{1}{2} w_{0} \rho D\int d \bhu  \int_{0}^{2 \pi} d \phi \int_{0}^{L} dt \left (t (\bhu \cdot \be_{i} )   \right )^{2} \nonumber \\
& \times \left [ (\ba \cdot {\bf v}_{0}(\phi) )^{2} - (\bn \cdot {\bf v}_{0}(\phi) )^{2}\right ] f_{U}(\bhu \cdot \bn)
\label{kanchor}
\end{align}
with $\be_{1} = \ba$, $\be_{2} = \bb$ and $\be_{3} = \bn$ as per the different modes. 
The integrals are easily solved with the aid of the Gaussian distribution by applying standard asymptotic expansion \cite{odijkoverview}. The resulting scaling expressions for the elastic modes take a simple form:
\begin{align}
K_{1}^{(s)} & \sim \frac{1}{4} w_{0} \frac{L^{2}}{D} \phi_{r}    \nonumber \\ 
K_{2}^{(s)} & \sim  \frac{1}{3} K_{1}^{(s)}    \nonumber \\
K_{3}^{(s)} & \sim   \frac{1}{9} w_{0} \frac{L^{2}}{D} \phi_{r} \frac{1}{\sigma}  
\label{ks}
\end{align}
where $\phi_{r} = (\pi/4)LD^{2}\rho$ denotes the rod volume fraction within the hybrid LC. 
We conclude that surface-anchoring between the rod inclusions and the molecular host primarily enhances the splay elasticity and,  to a lesser extent, the twist mode. The bend elasticity on the other hand seems hardly affected by the presence of the rod inclusions provided that the  anti-nematic order is strong ($\sigma \gg 1)$.

\subsection{Elasticity generated by rod correlations}

The elastic moduli associated with repulsive interactions between anti-nematically organized rods has been analyzed in detail by one of us in Ref. \cite{wensinkSM18}. The results are as  follows:
\begin{align}
K_{1}^{(r)} &   \sim \frac{k_{B}T}{D_{\rm eff}} \frac{ 1}{ \pi^{3}} (\phi_{r} \ell_{\rm eff})^{2} (\ln W + C_{1}) \nonumber \\ 
K_{2}^{(r)} & =   \frac{1}{3} K_{1}^{(r)}  \nonumber \\
K_{3}^{(r)} & \sim  \frac{k_{B}T}{D_{\rm eff}} \frac{4}{ 3 \pi^{3}} (\phi_{r} \ell_{\rm eff})^{2} \frac{ \ln W + C_{3}}{W}
\label{kranti}
\end{align}
with constants $C_{1} = \gamma_{E} -7/2 + \ln 24 \approx 0.255269$ and
$C_{3} =\gamma_{E} -23/6 + \ln 24 \approx -0.0780638$ and $\gamma_{E}$
 Euler's constant. The degree of anti-nematic orientation of the rods is expressed by an {\em
  effective} anti-nematic order parameter $ W = \sigma -  \frac{5}{4} \phi_{r} \ell_{\rm eff} S_{r}$ where $S_{r}$ denotes the conventional nematic order parameter of the rods which, in case of strong anti-nematic organization should be very close to its extreme value,  $S_{r} \approx -1/2$. 

The elastic constants further depend on the rod geometry  through the {\em effective} aspect ratio $\ell_{\rm eff} =(L/D)(D_{\rm eff}/D \gg \ell $ and rod diameter which roughly account for the electric double layers surrounding each rod.
Following Odijk {\em et al.} \cite{stroobants1986} we define:
\beq
D_{\rm eff} = D \left ( 1 + \frac{\ln A^{\prime} + \gamma_{E}  + \ln 2 + 1/2 }{\kappa D} \right ) 
\label{deff}
\eeq
with $\gamma_{E} \approx 0.577$ Euler's constant and $A^{\prime}$ an electrostatic amplitude given by (within the Debye-H\"{u}ckel approximation):
\beq
A^{\prime} \approx \frac{8 \pi v^{2} \ell_{B} e^{-\kappa D} } {\kappa^{3} D^{2} K_{1}^{2} (\kappa D/2)}
\label{aeff}
\eeq
with $v$ the effective line charge density defined as the number of elementary charges per unit rod length, $\kappa$ the Debye screening constant, $\ell_{B}$ the Bjerrum length and $K_{1}$ denotes a modified Bessel function (not to be confused with the splay elastic modulus).  The values relevant to our experiment are as follows; $v \approx 0.16e/nm$ which is sufficiently low  to justify the Debye-H\"{u}ckel approximation, $\kappa^{-1} \approx 120 nm$. This leads to $A^{\prime} \approx 105$ and an  effective-to-bare rod diameter $D_{\rm eff}/D \approx 28$.  

Similar to the surface anchoring contributions derived previously, the rod correlations primarily impact the splay modulus. This is in contrast to what is observed for conventional nematic order of rods where bend elasticity usually dominates \cite{odijkelastic}. This scenario we will encounter in Section III.  

 \subsection{Experimental measurement of the splay modulus}
 
\begin{figure}
\begin{center}
\includegraphics[width= \columnwidth ]{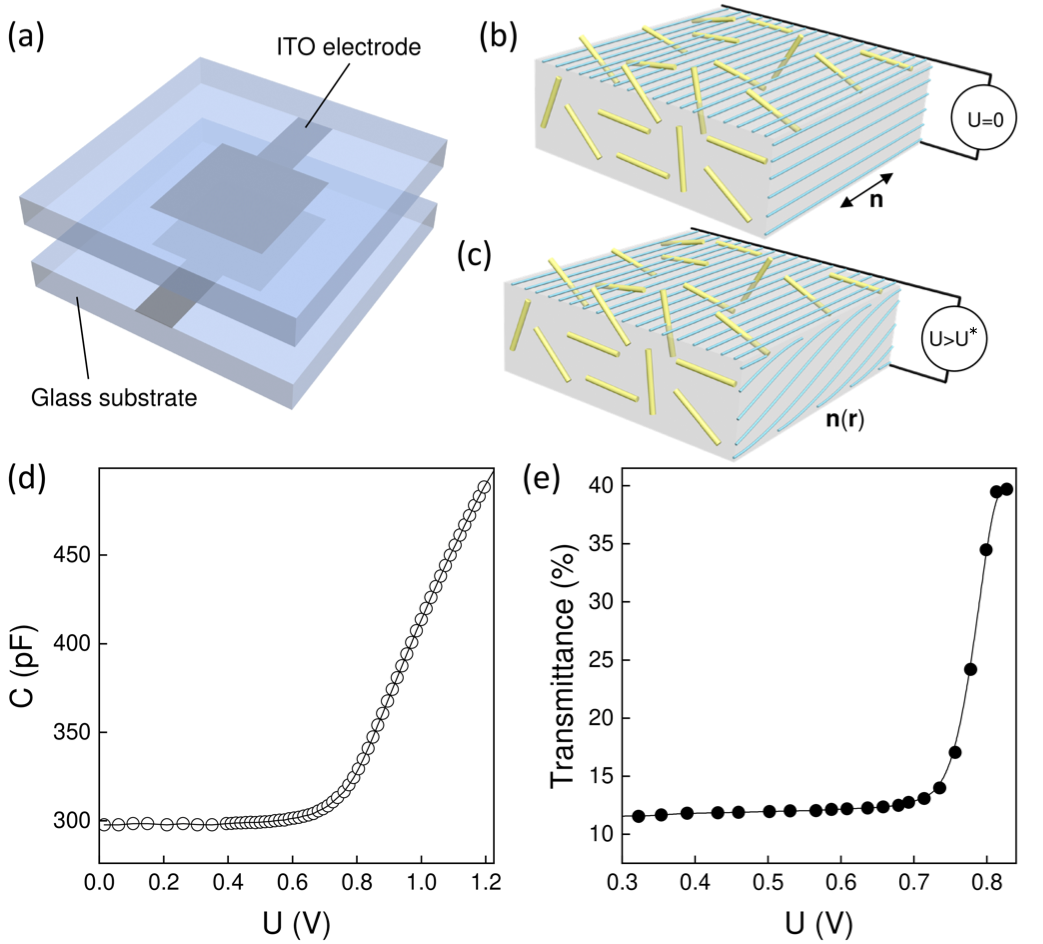}
\caption{ Measurements of the splay elastic constant $K_1$ of a hybrid LC: (a) Schematic diagram of a planar cell with patterned ITO electrodes. (b, c) A simplified schematic diagram of a hybrid LC slab without applied voltage (b) and above the threshold voltage $U^{\ast}$ (c) where the distorted molecular director $\bn({
\bf r)}$ is no longer spatially uniform.  The distortions are strongly exaggerated for illustrative purposes; in reality they are very weak at voltages just above the threshold value. (d) Capacitance and (e) transmittance of a hybrid LC in a planar cell as a function of the applied voltage. }
\label{exp2}
\end{center}
\end{figure}

Experimental measurements of the splay elastic constant $K_{1} $ were performed in a hybrid molecular-colloidal LC \cite{mundoor2018}, consisting of high-aspect-ratio inorganic colloidal nanorods dispersed in a nematic host [\fig{exp1}]. For our experiments, $\beta$-NaYF$_{4}$:Yb/Er nanorods [\fig{exp1}(a)] were synthesized by a hydrothermal synthesis method \cite{mundoor2018,synthesis2007}. They were treated using hydrochloric acid (HCl) to remove ligands used during synthesis and to control the length-to-diameter  ratio of the colloidal particles within $L/D \sim 40-110$ via slow etching of the solid nanocrystals. Hybrid nematics were prepared by re-dispersing nanorods in a commercially available pentylcyanobiphenyl (5CB, Frinton Labs, Inc.) nematic LC via solvent exchange and elevated-temperature evaporation, followed by cooling down the dispersion under vigorous stirring. Nematic LC dispersions of nanorods were infiltrated in between two glass substrates with patterned indium tin oxide (ITO) electrodes [\fig{exp2}(a)] spaced by glass microfibers setting the gap thickness to 20 $\mu$m, which was measured with a interferometric method. To achieve unidirectional planar boundary conditions for $\bn$, cell substrates were coated with 1wt.$\%$ aqueous polyvinyl alcohol or polyimide PI2555 (HD MicroSystem) and rubbed unidirectionally. We also used commercial planar cells purchased from Instec, Inc. Within the 5CB dispersions, the bare nanorod surface imparts weakly homeotropic  boundary conditions for $\bn$ [\fig{exp1}(b-f)]. We used small neodymium magnets (K$\&$J Magnetics, Inc.) to align nanorods as they orient perpendicular to the magnetic field lines \cite{mundoor2018}. A magnetic field applied to a sample was measured with a LakeShore 460 3-channel gaussmeter.

The splay elastic constant $K_1$ of a hybrid LC can be determined from measuring the threshold voltage $U^{\ast}$ of the Fr\'eedericksz transition for $\bn$ in a planar cell [\fig{exp2}(a-c)] upon the application of an electric field and using the relation $K_{1}=\pi^{-2} \epsilon_0 \Delta\epsilon (U^{\ast})^2 $, where $\epsilon_0$ is the dielectric permittivity in vacuum, $\Delta\epsilon = \epsilon_{\parallel}-\epsilon_{\perp}$ the dielectric anisotropy of a hybrid LC with $\epsilon_{\parallel}$ and $\epsilon_{\perp}$ the respective dielectric constants parallel and perpendicular to $\bn$ \cite{degennes1993}.

The threshold voltages $U^{\ast}$ were determined from the dependence of either a capacitance or optical transmittance of a planar hybrid LC cell on the applied voltage. The former [\fig{exp2}(d)] was measured at 1.0 kHz using an impedance gain-phase analyzer Schlumberger 1260 and the latter [\fig{exp2}(e)] was measured for a 632 nm laser beam passing through the cell placed between two crossed polarizers with $\bn$ oriented at $45^{\circ}$ to both polarizer and analyzer when an AC voltage (1.0 kHz) of continuously increasing amplitude above $U^{\ast}$ was applied.

 \begin{figure}
\begin{center}
\includegraphics[width= \columnwidth ]{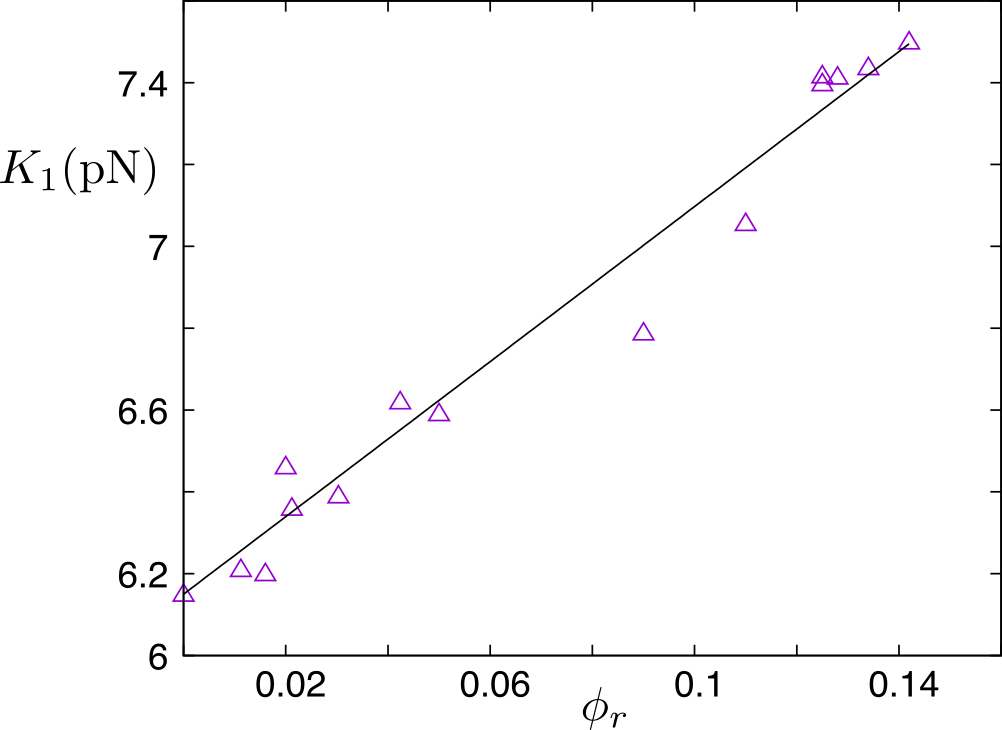}
\caption{ Splay elastic modulus $K_{1}$ of a hybrid LC depicted in \fig{exp1} as a function of the  rod volume fraction $\phi_{r}$ (in $\%$). The triangles denote experimental measurements, the solid line represent the (fit-free) theoretical prediction  according to \eq{ksuper}  which is dominated by  surface-anchoring mediated splay elasticity $K_{1}^{(s)}$ [cf. \eq{ks}]. The last three data points are located within the uniaxial-orthorhombic nematic coexistence region. }
\label{k1}
\end{center}
\end{figure}

The results for a range of rod concentrations are shown in \fig{k1}.  Taking the experimental values for  $K_{1}^{(0)} \sim 6.15 pN$ for pure 5CB, $\sigma \sim 188$, rod length $L \sim 1600 nm$, diameter $D \sim 25 nm$, dressed aspect ratio $\ell_{\rm eff} \sim 1536$, surface anchoring coefficient $w_{0}  =3.7.10^{-5} Jm^{-2}$, we predict from \eq{ks} that surface anchoring should enhance the splay elasticity by about $  0.94 pN$,  while the rod correlations give a negligibly small contribution  ($0.002 pN$, according to \eq{kranti}).  These predictions are corroborated by experimental results of the splay modulus as a function of the rod concentration. The linear increase with $\phi_{r}$ observed in \fig{k1} suggests  the enhancement of the splay elasticity being chiefly caused by surface-anchoring effects while rod-correlations play a marginal role.

\section{Uniaxial hybrid liquid crystals with co-aligned components}

For completeness, we also wish to address the case when surface anchoring forces the rods to co-align with the molecular director so that $\bm \parallel \bn$, as, for example, described in Ref. \cite{liu2014}. We may then repeat the analysis of Section II-A by imposing planar anchoring ${\bf v}_{0} = \bhu$ which forces the 5CB molecules to align along the principal rod direction. The surface anchoring energy per rod becomes:
\beq
U_{s} = -\frac{\pi}{2} DL w_{0} \cos^{2} \theta_{e} 
\eeq
which is  minimal at $\theta_{e} = 0$ or $\theta_{e} = \pi$ indicating that the rods are preferentially aligned along the molecular director $\bn$.
The orientational probability is given by a Boltzmann factor equivalent to \eq{fu}. For strong surface anchoring the probability approaches a simple Gaussian which reads:
\beq
f_{U}(\theta) \sim  \frac{\sigma}{4 \pi} \exp \left ( - \frac{\sigma}{2} \theta^{2}  \right )
\label{fuc}
\eeq
with amplitude:
\beq
\sigma = \frac{\pi LD w_{0}}{k_{B}T} \gg 1
\eeq
As before, knowledge of the rod orientation probability  enables us to analyze the two main elastic contributions, the one mediated by surface anchoring forces and the second one imparted by steric and electrostatic rod repulsion.

\subsection{Elasticity generated by surface anchoring }

The surface anchoring contribution can be readily estimated from the expression in \eq{kanchor}. Inserting the Gaussian probability \eq{fuc} we find in the limit of strongly aligned rods:
\begin{align}
K_{1}^{(s)} & \sim  \frac{4}{3\pi}k_{B}T \frac{L}{D^{2}} \phi_{r}    \nonumber \\ 
K_{2}^{(s)} & \sim  K_{1}^{(s)}    \nonumber \\
K_{3}^{(s)} & \sim   \frac{4}{3} w_{0} \frac{L^{2}}{D} \phi_{r}  
\label{ksco}
\end{align}
Clearly, since $w_{0} \gg k_{B}T$ and $L \gg D$ the bend modulus is affected the most, while the splay and twist contributions are identical and independent from the anchoring amplitude. If we assume a planar anchoring strength of about $w_{0} \sim 10^{-5} J/m^{2}$  and a concentration of $\phi_{r} \approx 0.1 \%$ we obtain  $K_{3}^{(s)} \approx 1 pN $, while the splay and twist counterparts are  at least about two orders of magnitude smaller.

\subsection{Elasticity generated by rod correlations}

In order to gauge the elastic resistance generated by rod interactions in a strongly ordered uniaxial nematic we simply quote the scaling results for the splay, twist and bend elasticity of infinitely thin rigid polyelectrolytes calculated by Vroege and Odijk \cite{odijkelastic, vroege1987}:   
\begin{align}
K_{1}^{(r)} &   \sim \frac{7}{8 \pi} (\phi_{r} \ell_{\rm eff}) \left [ 1 - \frac{1}{7} (1 + hY)^{-1} \right ]  \nonumber \\ 
K_{2}^{(r)} & \sim   \frac{1}{3}  K_{1}^{(r)}     \nonumber \\
K_{3}^{(r)}  & \sim  \frac{4}{3 \pi^{2}} (\phi_{r} \ell_{\rm eff})^{3} \left [ 1 + hY \right ]^{2}
\label{kvroege}
\end{align}
Noting that a stable nematic phase requires $\phi_{r}\ell_{\rm eff} \gg 1 $ we conclude that the bend modulus is much larger than the other two. As previously, $\ell_{\rm eff} \gg \ell$ denotes an effective aspect ratio correcting for the electric double layer repulsion among the rods. The effect of electrostatic twist, quantified by the parameter $h \equiv (\kappa D_{\rm eff})^{-1}$ has been analyzed in detail Ref. \cite{vroege1987}. The  twist effect relates to the fact that parallel rod configurations are strongly disfavored (with respect to perpendicular ones) because of the considerable overlap in electric double layers they entail \cite{stroobantslading}.  The factor $Y$ depends on the rod concentration and twist constant and follows from a transcendental equation:
\beq
Y = \ln [ \phi_{r} \ell_{\rm eff} (1+ hY)] -\frac{1}{2} \ln \pi + \frac{1}{2} \gamma_{E} - \frac{3}{2}
\eeq
which is easily solved numerically or through the use of iterative solutions proposed in Ref.  \cite{vroege1987}. In general, electrostatic repulsion leads to an enhanced increase with $\phi_{r}$ primarily for the bend modulus $K_{3}^{(r)}$ \cite{vroege1987}. To illustrate this, we consider a sample of $\phi_{r} =0.3 \%$ which, using the electrostatic parameters specified in II-B yields $h=0.18$ and $Y\approx 5$. The splay and bend moduli, respectively,  are then $K_{1}^{(r)} \approx 0.07 pN$ and $K_{3}^{(r)} \approx 0.28 pN$ which corresponds to a fairly large bend-splay ratio $K_{3}^{(r)} / K_{1}^{(r)} \approx 40$. 

Combining this with the predictions from \eq{ksco} we conclude  that creating a hybrid liquid crystal with co-aligned molecular and colloidal components provides an effective means to tune the bend-splay elastic ratio of the material. 

\section{Orthorhombic hybrid liquid crystals}

 \begin{figure}
\begin{center}
\includegraphics[width= \columnwidth ]{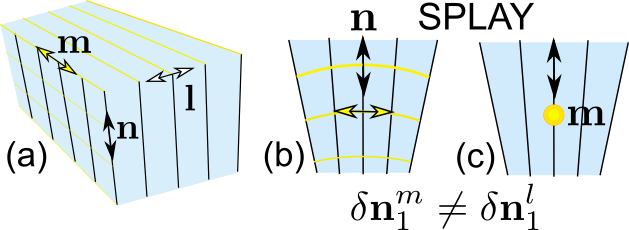}
\caption{  Nematic director frame in an orthorhombic nematic fluid with $D_{2h}$ point symmetry, spanned by a molecular director ($\bn$, black)  and  colloidal director ($\bm$, yellow). A uni-directional splay fluctuation $\delta \bn_{1}^{m}$ along $\bm$ generates bending of $\bm$ (splay-bend correspondence), which is absent for unidirectional splay  along $\bb$. Similar distinctions occur for the twist and bend fluctuations.  }
\label{ortho}
\end{center}
\end{figure}

We  now turn to the more complicated case of the  biaxial hybrid LC. This state  becomes stable at higher rod concentration where the rod correlations are strong enough to break the uniaxial symmetry in favour of an orthorhombic one characterized by mutually perpendicular molecular and colloidal directors $\bn \perp \ba$ [\fig{exp1}(e,f) and \fig{ortho}(a)].  
Govers and Vertogen \cite{govers1998} put forward a continuum theory for the  elasticity  of an orthorhombic biaxial nematic fluid which, in the absence of chiral interactions between the constituent molecules, involves twelve independent elastic constants. The deformation free energy is formally given by \cite{govers1998,stewartbiax}:
\begin{align}
& \frac{F_{el}}{V} = \frac{K_{1}}{2} (\nabla \cdot \bn)^{2}  +  \frac{K_{2}}{2} (\bn \cdot \nabla \times \bn)^{2} +   \frac{K_{3}}{2} (\bn \times \nabla \times \bn)^{2}   \nonumber \\ 
& +  \frac{K_{4}}{2} (\nabla \cdot \bm)^{2}  +  \frac{K_{4}}{2} (\bm \cdot \nabla \times \bm)^{2} +   \frac{K_{6}}{2} (\bm \times \nabla \times \bm)^{2}   \nonumber \\ 
& + \frac{K_{7}}{2} (\bn \cdot (\bm \times \nabla \times \bm))^{2}  +  \frac{K_{8}}{2}  (\bm \cdot (\bn \times \nabla \times \bn))^{2} \nonumber \\ 
&   + \frac{K_{9}}{2}  (\bm \cdot \nabla \times \bb)^{2}  +  \frac{K_{10}}{2}  (\bn \cdot \nabla \times \bb )^{2} +  \frac{K_{11}}{2}  (\nabla \times \bb )^{2} \nonumber \\
& +  \frac{K_{12}}{2}  (\nabla \cdot \bb )^{2}. 
\label{k12}
\end{align}
The uniaxial nematic elastic energy density, expressible in terms of the standard splay, twist and bend elastic constants, $K_{1}$, $K_{2}$ and $K_{3}$, respectively, can be recovered from the above expression by eliminating any contribution that contains the minor, colloidal director $\bm$.

\subsection{Classification of the orthorhombic elastic moduli}  

An intuitive way to rationalize the existence of twelve curvature elastic moduli for an orthorhombic nematic is to start from the consideration that in a biaxial nematic a unidirectional splay, twist or bend fluctuation each has two distinct directions given that the ${\mathcal O}(2)$ symmetry in the plane perpendicular to the molecular director $\bn$ is now broken.
Let us  recall from \eq{eldef1} that  ${\delta \bn}_{1}^{m}$ corresponds to a splay deformation of $\bn$ along the $\bm$-direction, ${\delta \ba}_{3}^{l}$ a unidirectional bend deformation of $\ba$ along the $\bb$ direction, and so forth. We are now in a situation where  director fluctuations $\delta \bn_{i}^{m}$  are no longer energetically equivalent to $\delta \bn_{i}^{l}$. This is illustrated in \fig{ortho}(b,c) for the splay distortion.  Based on this, we conjecture the presence of six independent elastic constants associated with deformations of $\bn$. Likewise, given that $\delta \bm_{i}^{n} \neq \delta \bm_{i}^{l}$ we must have six more independent moduli for the corresponding director fluctuations of $\bm$, giving a total of twelve.  Plugging in the parameterizations given by \eq{eldef1} into the continuum expression \eq{k12} we may straightforwardly identify each elastic modulus with a director specific deformation $\delta \bn$, $\delta \ba$ or a combination thereof. These are indicated in the second row of Table I.

The next step is to connect these fluctuations to the principal elastic moduli that we can attribute to being  generated either by pure 5CB, by rod correlations or by surface anchoring, as reflected by the superposition in \eq{ksuper}.
A few remarks are in order.
First we note that a splay-bend correspondence naturally emerges in orthorhombic biaxial systems since a splay deformation of one director induces a bend fluctuation along the perpendicular director and vice versa. This is illustrated in \fig{ortho}(b). 
Second, the surface-anchoring mediated elastic moduli are only relevant  when they involve a deformation of $\bn$ along $\bm$. The transversal ones along $\bb$ can safely be neglected as they probe ultraweak deviations of $\bn$ on the scale of the rod diameter.    By the same reasoning we will also ignore the effect of surface-anchoring on the director curvature of the colloids given that the curvature $\varepsilon$ will be much larger than the inverse rod length.  Taking these considerations into account we can readily identify the principal moduli associated with each $K_{i}$.  
\begin{table*}
\caption {Classification of  the twelve elastic moduli featuring in \eq{k12}  of an orthorhombic hybrid rod-molecular system in terms of the principal deformation modes of the individual components and the leading order contribution from the molecular host (denoted by superscript ``$0$"), surface-anchoring (``$s$") and rod-correlations (``$r$"). The values given in the last row are based on a hybrid LC with rod volume fraction of $\phi_{r} \sim 0.1 \%$. } \label{tab1} 
\begin{center}
{\setlength{\extrarowheight}{10pt}%
 \begin{tabular}{||c ||c | c |  c | c | c | c||} 
 \hline
 elastic modulus: & $  K_{1} \quad $ & $K_{2} \quad$ & $K_{3}\quad$ & $K_{4}\quad$ & $K_{5}\quad$ & $K_{6}\quad$  \\ 
 \hline 
 director-specific deformation: & $ {\delta \bn}_{1}^{l}  $ &  $ {\delta \bn}_{2}^{m}$ & ${\delta \bn }_{3}^{l}$ & ${\delta \ba}_{1}^{l}$ & ${\delta \ba }_{2}^{n}$  & ${\delta \ba}_{3}^{l}$ \\
 \hline 
principal modulus: &  $K_{1}^{(0)}$ & $K_{2}^{(0)} + K_{2}^{(s)}$ & $K_{3}^{(0)}$ & $K_{1}^{(r)}$ &$ K_{2}^{(r)}$ &$K_{3}^{(r)}$ \\
\hline
current experiment: &  $K_{1}^{(0)}$ & $K_{2}^{(0)}$ & $K_{3}^{(0)}$ & $\sim 0$ & $\sim 0$ &$K_{3}^{(r)}$ \\
\hline
estimated value (pN): &  6.15 & $  3 $ & $  10 $  & $ \sim 0$ & $ \sim 0$ & $ 0.3$ \\
 \hline
 \multicolumn{6}{c}{} \\
 \hline
elastic modulus: &   $  K_{7} \quad $ & $K_{8} \quad$ & $K_{9}\quad$ & $K_{10}\quad$ & $K_{11}\quad$ & $K_{12}\quad$  \\ 
  \hline
 director-specific deformation:  & $ {\delta \ba}_{3}^{n}$ &  ${\delta \bn}_{3}^{m}$ & ${\delta \bn}_{1}^{l}$ & ${\delta \ba}_{1}^{l}$ & ${\delta \bn }_{2}^{m} + {\delta \ba }_{2}^{n}$  & ${\delta \ba }_{3}^{l} + {\delta \bn }_{3}^{l}$ \\ 
 \hline
 principal modulus & $K_{3}^{(r)} + K_{1}^{(0)}$ & $K_{3}^{(0)} + K_{1}^{(r)} + K_{3}^{(s)} $ & $K_{1}^{(0)}$ & $K_{1}^{(r)}$ & $K_{2}^{(0)} +  K_{2}^{(r)} + K_{2}^{(s)} $ & $K_{3}^{(r)} + K_{3}^{(0)}$ \\
 \hline
 current experiment: & $K_{3}^{(r)} + K_{1}^{(0)}$ & $K_{3}^{(0)}  $ & $K_{1}^{(0)}$ & $\sim 0$ & $K_{2}^{(0)} $ & $K_{3}^{(0)} + K_{3}^{(r)}$ \\
 \hline
 estimated value (pN): &  6.45 & $  10 $ & $  6.15 $  & $ \sim 0$ & 3 & 10.3 \\
 \hline
 \end{tabular}}
\end{center}
\end{table*}
Comparing the upper two rows we find that the last four moduli can be linked to  (combinations of) the previous ones, namely:
\begin{align}
K_{9} & \sim K_{1} \nonumber \\
K_{10} & \sim K_{4} \nonumber \\
K_{11} & \sim K_{2} + K_{5} \nonumber \\
K_{12} & \sim K_{3} + K_{6} 
\end{align}
which implies that we only need to find scaling estimates for $K_{1}$ to  $K_{8}$. The leading-order contributions of the first six moduli may simply be associated with the  moduli of the respective pure components. These are indicated in the upper part of Table I. 

\subsection{Elasticity generated by surface anchoring}

We will now address the surface-anchoring contributions for a strongly biaxial hybrid LC. The  rod orientational probability  is assumed to be strongly peaked along  $\ba \perp \bn$. We write:
\begin{align}
f_{B}(\bhu ) & \sim \exp \left [ - \sigma \pp( \bhu \cdot \bn) + \beta ((\bhu \cdot \ba)^{2} - (\bhu \cdot \bb)^{2}) \right ] 
\end{align}
where the biaxial order parameter is assumed large ($\beta \gg 1$), in addition  to $\sigma \gg 1$ as for the uniaxial system (where $\beta =0$). Let us parameterize $\bhu \cdot \ba =  \cos \psi \sin \phi $, $ \bhu \cdot \bb = \cos \psi \cos \phi $ and $ \bhu \cdot \bn = \sin \psi $ in terms of the meridional angle $\psi$ and azimuthal angle $\phi$ with the molecular director $\bn$.  Expanding the argument for $\psi \gg 1$ we obtain the following asymptotic form: 
\begin{align}
f_{B}(\bhu ) & \sim \sqrt{\frac{3\sigma}{(2 \pi )^{3}  I_{0}^{2}(\beta)}}  \exp \left [ - \frac{3}{2} \sigma \psi^{2}  + \beta \cos 2 \varphi \right ] 
\end{align}
with $I_{0}$ a modified Bessel function. We reiterate that the surface-anchoring mediated moduli are defined as in \eq{kanchor} but with  $f_{U}$ replaced by the biaxial distribution above:
\begin{align}
K_{i}^{(s)} = &-\frac{1}{2} w_{0} \rho D\int d \bhu  \int_{0}^{2 \pi} d \phi \int_{0}^{L} dt \left (t (\bhu \cdot \be_{i} ) \right )^{2} \nonumber \\
& \left [ (\ba \cdot {\bf v}_{0}(\phi) )^{2} - (\bn \cdot {\bf v}_{0}(\phi) )^{2}\right ] f_{B}(\bhu ) 
\label{kanchor1}
\end{align}
with $\be_{1} = \ba$, $\be_{2} = \bb$ and $\be_{3} = \bn$ corresponding to splay, twist and bend, as before.  In the  limit of asymptotically strong rod alignment  along $\ba$  we evaluate the orientation average in the limit $\sigma \gg 1$ and $\beta \gg 1$.  Up to leading order for small $\psi$ we only find  only a nonzero contribution for the splay modulus:
\begin{align}
K_{1}^{(s)} & \sim \frac{2}{3} w_{0} \frac{L^{2}}{D} \phi_{r}  \nonumber \\
K_{2}^{(s)} & \sim 0  \nonumber \\
K_{3}^{(s)} & \sim 0
\label{ks1}
\end{align}
The fact that surface-anchoring leads to a much stronger  enhancement of the splay elasticity  than for the uniaxial nematic [cf. \eq{ks}] is not surprising because in the biaxial state the rods are strongly directed along $\bm$ where the impact of a splayed $\bn$ on the homeotropic surface anchoring is the largest (\fig{anchorfig}). Although $K_{1}^{(s)}$ can attain several pNs in magnitude, it does not feature in any of the elastic moduli listed in Table I. The surface anchoring elasticity, therefore, does not affect the nemato-elasticity of our orthorhombic hybrid nematic system, at least in the limiting case of strong rod alignment along $\ba$ we restrict ourselves to here.  

\subsection{Elasticity generated by rod correlations}

The formation of a stable orthorhombic nematic fluid requires elevated rod concentration where the moduli associated with  nanorod correlations $K_{j}^{(r)}$ ($j=1,2,3$) are expected to be much larger than those for the relatively dilute uniaxial nematic.  In order to estimate the extent to which rod interactions dominate the elastic properties of the hybrid LC  we  use the scaling predictions shown previously in \eq{kvroege}. We infer that at the highest rod concentration probed in experiment $\phi_{r} =0.142 \%$, the bend elasticity generated by  the charged rods is much smaller than that of 5CB ($K_{3}^{(0)} \approx 10 pN$) so it seems justified to ignore all contributions in \eq{k12} that involve splayed and twisted distortions of $\ba$. 

\subsection{Leading order moduli for an orthorhombic hybrid nematic}

Having demonstrated that both surface-anchoring terms are much weaker than those due to rod-correlations and noting that the rod-generated splay and twist elastic moduli are negligible compared to the dominant bend modulus, we arrive at the following leading order estimates:
\begin{align}
K_{7} &\sim K_{1}^{(0)} + K_{3}^{(r)} \nonumber \\
K_{8} &\sim K_{3}^{(0)} 
\end{align}
Applying the same approximations to all twelve constants featuring in the continuum theory \eq{k12} we arrive at a much more manageable set of moduli that only depend on the known values for pure 5CB and the bend elastic constant of the immersed rods $K_{3}^{(r)}$. We wish to underline that the estimates only make sense for the current hybrid molecular-rod nematic system which consists of slender rods with particular combination of electrochemical properties regarding surface charge and screening. For instance, the balance between surface anchoring and intercolloidal  forces is likely to be quite different for short rods for which surface anchoring contributions play a  more prominent role. Also, the bend-splay elastic anisotropy for conventional nematic order is expected to be different in view of the intricate electrostatic interactions between finite-aspect-ratio colloidal particles \cite{eggen2009,everts2020}. We  reiterate that all predictions presented here are subject to the condition that the rods be strongly aligned along their director $\ba$.

Going back to the experimental system at hand and recalling that the rod-driven bend modulus $K_{3}^{(r)} \approx 0.28 pN$ is still an order of magnitude smaller than the smallest elastic modulus of pure 5CB ($K_{2}^{(0)} \approx 3 pN$) we may even contemplate a more stringent reduction of the biaxial moduli by retaining only the contributions from the molecular host. A minimal continuum expression  can be obtained by applying the commonly used one-constant approximation $K_{1}^{(0)} \equiv K_{2}^{(0)} \equiv K_{3}^{(0)} = K$. Using basic vector manipulations based on   Lagrange's identity $| {\bf a} \times {\bf b} |^{2} = |{\bf a} |^{2} | {\bf b}|^{2} - ({\bf a} \cdot {\bf b} )^{2}$ and the triple vector identity ${\bf a} \cdot ({\bf b} \times {\bf c}) = {\bf b} \cdot ({\bf c} \times {\bf a})$ we find the following (strongly) simplified expression for the
Frank elastic free energy for our hybrid biaxial nematic:
\begin{align}
& \frac{F_{el}}{V}  \approx \frac{K}{2} \left [ (\nabla \cdot \bn)^{2}  + (\nabla \cdot \bb)^{2} + | \nabla \times \bn |^{2} + | \nabla \times \bb |^{2} \right . \nonumber \\
& \left . + \{\ba \cdot (\bn \times (\nabla \times \bm))\}^{2} +  \{\ba \cdot (\bn \times (\nabla \times \bn))\}^{2} \right . \nonumber \\
& \left . + (\ba \cdot (\nabla \times \bb))^{2} \right ]
\label{felsimple}
\end{align}
We will not attempt to further simplify this expression. An interesting feature of \eq{felsimple} is that even though the expression should be applicable only to hybrid LCs whose rod correlations are pronounced enough to enforce strong alignment of the colloidal component along $\bm$, the relevant elastic modulus is chiefly governed by elastic forces generated by the molecular LC alone. This is consistent with the main conclusion of the density-functional study of  Ref. \cite{longa2} where the elastic fluctuations of the minor director (in this case $\ba$) were found  to play a minor role

\section{Discussion}

\begin{figure}
\begin{center}
\includegraphics[width= \columnwidth ]{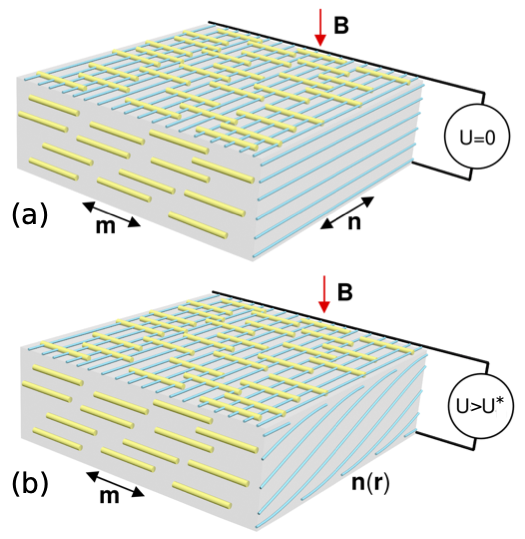}
\caption{ Schematic overview of an electric-field-induced Fr\'{e}edericksz  transition within an orthorhombic biaxial nematic LC in a planar cell. (a) Without applied voltage ($U=0$).   (b)  If the  voltage applied across the slab  exceeds the threshold value $U^{\ast}$  a splay deformation of the molecular director $\bn$ is induced while simultaneous realignment of the colloidal rod director $\ba$ is prevented by applying a weak auxiliary magnetic field ${\bf B}$ .   }
\label{exp3}
\end{center}
\end{figure}

The interrelation between the orthorhombic elastic moduli  as borne out from our scaling theory, although based on solid theoretical arguments,  remains largely speculative and the predictions evidently call for further experimental validation. However, the use of the conventional method of electro-magnetic-field induced director distortions (Fr\'{e}edricksz transition), as  we did for the uniaxial case in Section II-C, poses a number of technical complications that are specific to these low-symmetry orthorhombic hybrid LCs. The most important ones are the following:

\begin{itemize}

\item  Identifying  all twelve elastic moduli that feature in \eq{k12} requires a considerable variety of different LC cells with specific boundary conditions for each component as well as external field directions. A tentative strategy to extract the orthorhombic elastic moduli from specific LC set-ups is discussed in the Appendix. 

\item Both molecular and colloidal directors may be distorted at similar external field strengths, which makes it hard to disentangle specific director deformation of one component, while keeping the other one unaffected. This is illustrated in \fig{exp3} for the case of an electric-field generated splay distortion  within the molecular director field $\bn$ (corresponding to $K_{1}$ in Table I). In this particular case,  simultaneous distortions of the colloidal director $\bm$ can be suppressed by applying a weak additional magnetic field where the negative magnetic anisotropy of the colloidal rods forces the rods to align  perpendicular to ${\bf B}$. We note that without such an auxiliary external magnetic field, both directors tend to respond simultaneously to the applied electric field with the colloidal director responding at even lower voltages than the molecular LC,  making it difficult to measure the elastic constants associated with deformations of the molecular director alone.  

\item The measurements require full control of the anchoring conditions for {\em both} molecular and colloidal directors. While strong planar and homeotropic anchoring at the cell walls is relatively straightforward to achieve for the molecular LC director using standard techniques used for conventional thermotropic LCs, controlling the anchoring of colloidal rods within such hybrid systems is far from trivial. The methods to impose strong tangential or homeotropic anchoring on the colloidal director needed for such experiments still have to be developed (so far only weak surface anchoring boundary conditions for the colloidal director have been demonstrated).

\item In an orthorhombic hybrid LC the nematic order  of  both components measured with reference to their respective principal directors is biaxial \cite{mundoor2018}. Consequently, the dielectric (and diamagnetic) tensor of each component alone becomes biaxial too, now featuring three different principal values of dielectric (and diamagnetic) constants.  For example, the magnetic energy density of a system with orthorhombic symmetry  subject to an applied magnetic field ${\bf H}$ formally reads \cite{stewartbiax}:
\beq
\frac{F_{m}}{V} = -\frac{\mu_{0}}{2}  [ \chi_{nl} ({\bf H} \cdot \bn )^{2} + \chi_{ml}({\bf H} \cdot \ba )^{2} + \chi_{l} H^{2}   ] \nonumber
\eeq
where $\mu_{0}$ denotes the vacuum permeability and $\chi_{nl}  = \chi_{n} - \chi_{l} $ and $\chi_{ml}  = \chi_{m} - \chi_{l} $  the two relevant diamagnetic susceptibility anisotropies with respect to the orthorhombic director tripod (\fig{ortho}a).  Analogously, in case of an applied electric field ${
\bf E}$ the electric energy density reads:
\beq
\frac{F_{e}}{V} = -\frac{\epsilon_{0}}{2} [ \epsilon_{nl} ({\bf E} \cdot \bn )^{2} + \epsilon_{ml}({\bf E} \cdot \ba )^{2} + \epsilon_{l} E^{2}   ] \nonumber
\eeq
with $\epsilon_{0}$ the vacuum permittivity and $\epsilon_{ij}  = \epsilon_{i} - \epsilon_{j} $  the dielectric permittivity anisotropy. The experimental determination of the respective dielectric and diamagnetic anisotropies, from which the response of the molecular and colloidal directors to the external electric or magnetic fields can be assessed, adds another level of complexity to measuring the elastic properties of orthorhombic hybrid LCs. In the Appendix we show that, as a first approximation, the diamagnetic energy density of an orthorhombic system can be written in terms of a superposition of two uniaxial nematic materials with mutually perpendicular principal alignment axes.
Even in a uniaxial hybrid LC the dielectric response of the molecular LC is affected by the presence of the colloidal rods, although this effect can be straightforwardly accounted for through direct measurements, as was done in this study (Section II-C).  

\end{itemize}
It is clear that these  technical complications  impede a straightforward extension of the measurements outlined in Section II-C towards the orthorhombic case, at least when using techniques based on measurements of the realignment thresholds commonly utilized for  uniaxial nematics. Alternative methods to address the nemato-elastic response of liquid crystals include, for instance, those based on light scattering \cite{degennes1993,borsali1998,kimmich2012} where thermal fluctuations of  the nematic director are probed directly, without the need to apply an external field. In this case, however, a theoretical framework needs to be developed to establish delicate relationships between light scattering observables under various polarization and geometric alignment conditions and the different elastic constants of interest. 

\section{Conclusions}

Inspired by recent experimental studies of strongly anisotropic colloidal rods immersed in thermotropic 5CB, we have presented a theoretical analysis of the nematic elastic moduli of such hybrid LCs, starting from considerations of surface anchoring and correlations at the level of the individual rods.   Two classes of nematic symmetries are considered; uniaxial systems in which the rods are organized (anti-)nematically within the molecular LC, and  a biaxial system in which rod-rod interactions are strong enough to stabilize orthorhombic order characterized by each component aligning along mutually perpendicular directors. 

We find that in the uniaxial state the elasticity of the molecular LC is enhanced primarily by surface anchoring effects  with the splay modulus increasing linearly with the concentration of rods, while the twist and bend moduli remain virtually unaffected by the presence of the rods.   The enhanced splay mode as predicted by theory is corroborated by experimental results, both demonstrating that $K_{1}$ of pure 5CB can be increased by about 20 \% for rod volume fractions as low as 0.1 \%.  

Having gained confidence in the model, we proceed to a theoretical analysis of the elastic moduli for the orthorhombic hybrid LC.  We classify the twelve elastic constants that feature in the generalized Oseen-Frank expression for biaxial nematics \cite{govers1998} in terms of principal contributions which may stem from the LC solvent alone, from surface anchoring between the rod surface and the solvent, or from steric and electrostatics-driven interactions between the immersed rods. This enables us to identify and classify the principal elastic moduli in our hybrid LC in terms of the known bulk moduli of the LC solvent, while the surface-anchoring and correlation-mediated contributions are predicted by theory. Explicit values for these principal moduli are given that fully specify the elastic anisotropy of our hybrid molecular-colloidal LC.  A compact expression for the corresponding elastic  energy then follows from a simple one-constant approximation in which the moduli pertaining the LC solvent are assumed equal. 
Despite its simplified form, the proposed elastic energy of our hybrid biaxial nematic LC remains highly non-trivial and is expected to generate complex director topologies that are fundamentally different from their uniaxial counterparts.  We briefly discuss the experimental difficulties that arise when attempting to measure the elastic constants of an orthorhombic nematic material using threshold field strengths for director switching based on conventional LC cells, and provide a possible roadmap towards systematically extracting all relevant moduli using  Fr\'{e}edericksz transitions \cite{degennes1993}.     

The results of this work might inspire future experimental and modelling studies exploring the complex nemato-elastic properties of hybrid molecular-colloidal LCs as a new breed of orthorhombic materials combining fluidity with low-symmetry orientational order \cite{mundoor2018}. These studies may eventually pave the way to exploring  ``biaxial-nematic colloids", where the broken uniaxial symmetry of, for instance, a spherically symmetric colloidal inclusion imparted by its orthorhombic environment as well as the ensuing anisotropic colloid-colloid interactions may be harnessed to explore even more complex manifestations of topological colloids \cite{senyukh2013} or hierarchical self-assembly than have thus far been studied for uniaxial nematic colloids. 

\begin{acknowledgments}

We thank B. Fleury, Q. Liu, L. Longa, J.-S. Tai and J.-S. Wu for discussions. We acknowledge support by the U.S. Department of Energy, Office of Basic Energy Sciences, Division of Materials Sciences and Engineering, under contract DE-SC0019293 with the University of Colorado at Boulder (B.S., H.M. and I.I.S.).

\end{acknowledgments}

\section*{Appendix:  Roadmap to measuring the orthorhombic elastic moduli using Fr\'{e}edericksz transitions}

\begin{figure}
\begin{center}
\includegraphics[width= 0.9\columnwidth ]{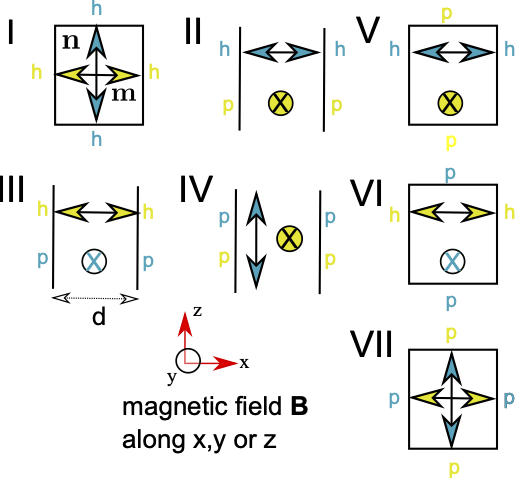}
\caption{ Sketch of possible LC cells that would need to be realized to measure all twelve elastic moduli for a biaxial hybrid molecular-colloidal LC.  In all cases, strong anchoring is assumed of the molecular director ($\bn$) and the colloidal one ($\bm$). For each director the anchoring is fixed along a single set of opposing walls with the same anchoring type [homeotropic (h) or planar (p)]. Combining the different anchoring scenarios and the three orthogonal magnetic field directions one obtains a large range of different configurations denoted by ${\rm I}_{x}, {\rm I}_{y}, {\rm I}_{z}$, et cetera.  }
\label{cells}
\end{center}
\end{figure}

In Section II-C we discussed  measurements of the splay modulus of a uniaxial hybrid LC from locating the corresponding Fr\'eedericksz transition. Here, we explore the full range of transitions that could potentially be realized by varying the anchoring conditions for each component as well as the direction of the magnetic field.  In \fig{cells} we have sketched a number of different LC cells one could envisage for a hybrid molecular-colloidal LC.  As for conventional uniaxial systems \cite{degennes1993}, the Fr\'eedericksz transition associated with each particular set-up  enables us to probe a particular elastic modulus (or a combination of several moduli) of the orthorhombic material that we will explore below. 

We start from the  
magnetization  {\bf M}  of a single uniaxial rod (of either molecular or colloidal origin) with orientation $\bn$  characterized by a parallel diamagnetic susceptibility  ($\chi_{\parallel}$) and a perpendicular one ($\chi_{\perp}$) with respect to the principal rod axis  in response to  an applied magnetic field {\bf H} \footnote{An analogous description follows for the case of an applied electric field ${
\bf E}$ acting on a uniaxial rod with dielectric permittivity $\epsilon_{\parallel}$ and $\epsilon_{\perp}$.}: 
\beq
{\bf M} = \mu_{0} [ \chi_{\parallel} (\bn \otimes \bn \cdot {\bf H} ) + \chi_{\perp} ({\bf I} - \bn \otimes \bn) \cdot {\bf H} ]
\eeq
with $\otimes$ denoting a dyadic product and $\mu_{0}$ the vacuum permeability. The magnetic energy per rod is given by:
  \beq
  F_{ m}^{\rm rod} =  -\frac{1}{2} ({\bf H} \cdot {\bf M}) = -\frac{1}{2} \mu_{0}^{-1} [\Delta \chi ({\bf B} \cdot \bn)^{2} + \chi_{\perp} B^{2}]
  \eeq
  in terms of the magnetic induction ${\bf B} = \mu_{0} ({\bf H} + {\bf M}) \approx \mu_{0} {\bf H} $ for weak magnetic susceptibility. The last term is immaterial for the present analysis while $\Delta \chi = \chi_{\parallel} - \chi_{\perp} $ denotes the susceptibility anisotropy of the rod. For our hybrid LC  we  express the magnetic energy in terms of a superposition of the molecular and colloidal components both assumed to be uniaxially aligned along their respective directors. For notational convenience  we set the vacuum permeability $\mu_{0}$ to unity without loss of generality. The magnetic energy of the orthorhombic hybrid system then reads:
 \begin{align}
 F_{ m} &\sim  -\frac{1}{2}  \int dV  [  \Delta \chi^{(0)} ({\bf B} \cdot \bn)^{2} + \Delta \chi^{(r)} ({\bf B} \cdot \bm)^{2} ]
 \end{align}
 so that
 \begin{align}
\frac{ F_{ m} }{V}&\sim  -\frac{1}{2} \phi_{0} \Delta \chi^{(0)} ({\bf B} \cdot \bn)^{2} - \frac{1}{2} \phi_{r} \Delta \chi^{(r)} ({\bf B} \cdot \bm)^{2} 
\label{fmag}
 \end{align}
 with $\phi_{0/r}$  the volume fraction of each component (where generally  $\phi_{0} \gg \phi_{r}$) and $\Delta \chi^{(0/r)}$ the  respective diamagnetic susceptibility anisotropies  which may differ in amplitude and even sign.
 The director field distorted by the  magnetic field can be parameterized in terms of two independent angles $\theta$ and $\varphi$ denoting spatially varying coupled distortions of the molecular and rod directors. For setup ${\rm I}_{x}$ we write: 
 \begin{align}
 \bn &= (\sin \theta(z), 0 , \cos \theta(z)), \nonumber \\
 \bm &= (\cos \varphi(x), 0 , \sin \varphi(x)),\nonumber \\ 
 {\bf B} & = B(1,0,0) 
 \end{align}
 All other cases can be parameterized likewise.

 Inserting all the parameterizations back into the Oseen-Frank elastic  (\eq{k12}) and magnetic free energy (\eq{fmag}) we obtain  expressions for the total free energy change per unit volume.  The rest of the analysis proceeds in a  way similar to that of the conventional (uniaxial) Fr\'edericksz transition. A formal minimization of the free energy  with respect to $\theta$ and $\varphi$ yields a set of coupled Euler-Lagrange equations  that can be linearized for small angular fluctuations. Keeping only the linear order terms we obtain two {\em decoupled} second-order differential equations. For instance, for the particular cell geometry ${\rm I}_{x}$ these read:
  \begin{align}
 B^{2} \phi_{0} \Delta \chi^{(0)}  \theta(x) &= - (K_{3}  + K_{8}) \theta''(x)  \nonumber \\
B^{2}  \phi_{r} \Delta \chi^{(r)} \varphi(x) & = (K_{6} + K_{7}) \varphi ''(x) 
  \end{align}
 Substituting $\theta(x) = \theta_{0} e^{i q x}$ and  $\varphi(x) = \varphi_{0} e^{ik x}$ enables us to derive  expressions for the threshold amplitude $B^{\ast}$. 
 The boundary conditions require that $q,k=\pi/d$ in terms of the wall-to-wall distance $d$. Keeping only the real contributions for $B^{\ast}$ we obtain:   
    \begin{align}
 {\rm I}_{x}: \hspace{0.1cm} B^{\ast} &= q \sqrt{ \frac{K_{3} + K_{8} }{\phi_{0}  \Delta \chi^{(0)}}} \nonumber \\
 {\rm I}_{y}: \hspace{0.1cm} B^{\ast} &= k \sqrt{ \frac{K_{6} + K_{12} }{\phi_{r}  \Delta \chi^{(r)}}}  \nonumber \\
 {\rm I}_{z}: \hspace{0.1cm} B^{\ast} &= k \sqrt{ \frac{K_{6} + K_{7} }{\phi_{r}  \Delta \chi^{(r)}}}  
 \label{axyz}
  \end{align} 
  and similarly for cell geometry II:
    \begin{align}
 {\rm II}_{x}: \hspace{0.1cm} B^{\ast} &= k \sqrt{ \frac{K_{4}  }{\phi_{r}  \Delta \chi^{(r)}}} \nonumber \\
 {\rm II}_{y}: \hspace{0.1cm} B^{\ast} &= q \sqrt{ \frac{K_{3} + K_{8} }{\phi_{0}  \Delta \chi^{(0)}}}  \nonumber \\
 {\rm II}_{z}: \hspace{0.1cm} B^{\ast} &= \begin{cases} q \sqrt{ \frac{K_{3} + K_{12} }{\phi_{0}  \Delta \chi^{(0)}}} \\
 k \sqrt{ \frac{K_{5} + K_{11} }{\phi_{r}  \Delta \chi^{(r)}}}
 \end{cases}
  \end{align}   
The threshold magnetic field for all other set-ups are then as follows: 
    \begin{align}
 {\rm III}_{x}: \hspace{0.1cm} B^{\ast} &= q \sqrt{ \frac{K_{1}  }{\phi_{0}  \Delta \chi^{(0)}}} \nonumber \\
 {\rm III}_{y}: \hspace{0.1cm} B^{\ast} &= k \sqrt{ \frac{K_{6} + K_{7} }{\phi_{r}  \Delta \chi^{(r)}}}  \nonumber \\
 {\rm III}_{z}: \hspace{0.1cm} B^{\ast} &= \begin{cases} q \sqrt{ \frac{K_{2} + K_{11} }{\phi_{n}  \Delta \chi^{(n)}}} \\
 k \sqrt{ \frac{K_{6} + K_{12} }{\phi_{r}  \Delta \chi^{(r)}}} 
 \end{cases}
  \end{align}    
  
    \begin{align}
 {\rm IV}_{x}: \hspace{0.1cm} B^{\ast} &= \begin{cases} q \sqrt{ \frac{K_{1}  + K_{9} + K_{11}  }{\phi_{0}  \Delta \chi^{(0)}}} \\
 k \sqrt{ \frac{K_{4}  + K_{10} + K_{11}  }{\phi_{r}  \Delta \chi^{rm)}}} \end{cases} \nonumber \\
 {\rm IV}_{y}: \hspace{0.1cm} B^{\ast} &= q \sqrt{ \frac{K_{2} }{\phi_{0}  \Delta \chi^{(0)}}}  \nonumber \\
 {\rm IV}_{z}: \hspace{0.1cm}  B^{\ast} &= k \sqrt{ \frac{K_{5}  }{\phi_{r}  \Delta \chi^{(r)}}}  
  \end{align}    
  
      \begin{align}
 {\rm V}_{x}: \hspace{0.1cm} B^{\ast} &= k \sqrt{ \frac{K_{5}  }{\phi_{r}  \Delta \chi^{(r)}}} \nonumber \\
 {\rm V}_{y}: \hspace{0.1cm} B^{\ast} &= q \sqrt{ \frac{K_{3} + K_{8} }{\phi_{0}  \Delta \chi^{(0)}}}  \nonumber \\
 {\rm V}_{z}: \hspace{0.1cm} B^{\ast} &= \begin{cases} q \sqrt{ \frac{K_{3} + K_{12} }{\phi_{0}  \Delta \chi^{(0)}}} \\ k \sqrt{ \frac{K_{4}  + K_{10} + K_{11} }{\phi_{r}  \Delta \chi^{(r)}}}
 \end{cases}
  \end{align}   
  
      \begin{align}
 {\rm VI}_{x}: \hspace{0.1cm} B^{\ast} &= q \sqrt{ \frac{K_{2}  }{\phi_{0}  \Delta \chi^{(0)}}} \nonumber \\
 {\rm VI}_{y}: \hspace{0.1cm} B^{\ast} &= k \sqrt{ \frac{K_{6} + K_{7} }{\phi_{r}  \Delta \chi^{(r)}}}  \nonumber \\
 {\rm VI}_{z}: \hspace{0.1cm} B^{\ast} &= \begin{cases} q \sqrt{ \frac{K_{1} + K_{9} + K_{11} }{\phi_{0}  \Delta \chi^{(0)}}} \\
 k \sqrt{ \frac{K_{5}  + K_{6} + K_{12} }{\phi_{r}  \Delta \chi^{(r)}}}  
 \end{cases}
  \end{align}  
  and
      \begin{align}
 {\rm VII}_{x}: \hspace{0.1cm} B^{\ast} &= q \sqrt{ \frac{K_{1}  }{\phi_{0}  \Delta \chi^{(0)}}} \nonumber \\
 {\rm VII}_{y}: \hspace{0.1cm} B^{\ast} &= \begin{cases} q \sqrt{ \frac{K_{2} + K_{11} }{\phi_{0}  \Delta \chi^{(0)}}} \\
 k \sqrt{ \frac{ K_{11} }{\phi_{r}  \Delta \chi^{(r)}}} \end{cases} \nonumber \\
  {\rm VII}_{z}: \hspace{0.1cm} B^{\ast} &= k \sqrt{ \frac{K_{4} }{\phi_{r}  \Delta \chi^{(r)}}}  
  \end{align}  
 Wherever two  expression are given, only the one giving the lowest threshold magnetic field $B^{\ast}$ will be of physical significance.  From the set-ups described thus far we are  able to identify the following six moduli:
      \begin{align}
      {\rm III}_{x}  &\rightarrow K_{1} \nonumber \\
      {\rm IV}_{y} &\rightarrow  K_{2} \nonumber \\
      {\rm II}_{x} & \rightarrow K_{4} \nonumber \\
      {\rm IV}_{z} & \rightarrow K_{5} \nonumber \\
      {\rm III}_{z} & \rightarrow K_{11} \nonumber \\
      {\rm IV}_{x} \; {\rm or} \; {\rm VI}_{z} & \rightarrow K_{9}
  \end{align}  
 Furthermore, if one could design a set-up in which the Fr\'eedericksz transition of $\bm$  precedes that of $\bn$,  one could extract $K_{10}$ via
 \beq
 {\rm V}_{z} \; {\rm or} \; {\rm IV}_{x} \rightarrow K_{10}
 \eeq
 The last five modes can be obtained as follows. First, upon close inspection of \eq{k12} one deduces that $K_{7}$ and $K_{8}$ represent a bend deformation of $\bm$ specifically along the direction $\bn$ and a bend deformation of $\bn$ along $\bm$, respectively.  The corresponding  bend moduli for the {\em uniaxial} systems $K_{6}$ and $K_{3}$ are, as required by symmetry, invariant with respect to the direction along which a bend deformation is applied.   The following four moduli can thus be determined from (there are several possible set-ups):
      \begin{align}
      {\rm II}_{y} \rightarrow K_{3} = K_{8} \nonumber \\
      {\rm III}_{y} \rightarrow K_{6} = K_{7} 
  \end{align}  
 which finally leaves us with $K_{12}$ (here also several set-ups are possible):
 \beq
 {\rm II}_{z} \rightarrow K_{12}.
  \eeq
with which all twelve elastic moduli have been identified. 
 We wish to underline that the schematic outlined above is by no means unique and that other, possibly more  experimentally viable strategies may be conceivable based on different combinations of cell geometry and anchoring conditions, as well as both electric and magnetic field application, and even combinations of the two. At the same time, the most significant challenge for experimentally measuring the complete set of elastic constants of hybrid nematic LCs based on realignment thresholds relates to the need of defining strong homeotropic and planar anchoring for the colloidal director, for which the suitable methodologies still need to be developed. A different approach could involve elastic constant measurements based on light scattering for different polarization and director orientation geometries, which is also worth considering.

\bibliographystyle{apsrev}
\bibliography{refs}

\end{document}